\documentclass[aps,pra,twocolumn,showpacs,floatfix,superscriptaddress,reprint,
nobalancelastpage]{revtex4-1}
\usepackage[pdftex]{color}
\usepackage{hyperref}
\usepackage[english]{babel}
\usepackage{amsfonts}
\usepackage{amsmath}
\usepackage{graphicx}
\usepackage[caption=false]{subfig}
\usepackage[modulo]{lineno}

\usepackage{soul}
\usepackage{color}
\usepackage{ulem}

\newcommand{\eps}{\epsilon}
\newcommand{\expo}[1]{\mathrm{exp}\left(#1\right)}

\newcommand{\cose}[1]{\mathrm{cos}\left(#1\right)}

\newcommand{\seno}[1]{\mathrm{sin}\left(#1\right)}

\newcommand{\ex}[1]{\mathrm{e}^{#1}}
\newcommand{\Ket}[1]{\left|#1\right>}
\newcommand{\Bra}[1]{\left<#1\right|}
\newcommand{\BraKet}[2]{\left<#1|#2\right>}
\newcommand{\KetBra}[2]{|#1\rangle\langle#2|}

\begin{document}

\title{Time-optimal control fields for quantum systems with multiple avoided crossings}

\author{P. M. Poggi}
\email{ppoggi@df.uba.ar}
\affiliation{Departamento de F\'{\i}sica Juan Jose Giambiagi 
 and IFIBA CONICET-UBA,
 Facultad de Ciencias Exactas y Naturales, Ciudad Universitaria, 
 Pabell\'on 1, 1428 Buenos Aires, Argentina}
\author{F. C. Lombardo}
\affiliation{Departamento de F\'{\i}sica Juan Jose Giambiagi 
 and IFIBA CONICET-UBA,
 Facultad de Ciencias Exactas y Naturales, Ciudad Universitaria, 
 Pabell\'on 1, 1428 Buenos Aires, Argentina}
\author{D. A. Wisniacki}
\affiliation{Departamento de F\'{\i}sica Juan Jose Giambiagi 
 and IFIBA CONICET-UBA,
 Facultad de Ciencias Exactas y Naturales, Ciudad Universitaria, 
 Pabell\'on 1, 1428 Buenos Aires, Argentina}

\date{\today}

\begin{abstract}
We study time-optimal protocols for controlling quantum systems which show several avoided level crossings in their energy spectrum. The structure of the spectrum allows us to generate a robust guess which is time-optimal at each crossing. We correct the field applying optimal control techniques in order to find the minimal evolution or quantum speed limit (QSL) time. We investigate its dependence as a function of the system parameters and show that it gets proportionally smaller to the well-known two-level case as the dimension of the system increases. Working at the QSL, we study the control fields derived from the optimization procedure, and show that they present a very simple shape, which can be described by a few parameters. Based on this result, we propose a simple expression for the control field, and show that the full time-evolution of the control problem can be analytically solved. 
\end{abstract}

\maketitle

\section{Introduction}

The development of future communication and information technologies is expected to rely heavily on the precise manipulation of physical systems at the nano- and sub-nanoscale. For this reason, coherent control of quantum systems has become a major goal in physical sciences over the past decades. In this context, the design and implementation of quantum control methods has raised a lot of interest, and many theoretical \cite{bib:shapiro2011,bib:wiseman2009,bib:dalessandro2008} and experimental \cite{bib:meshulach1998,bib:press2008} works have been devoted to this subject.\\

Controlled quantum operations are tipically intended to be performed in the fastest possible way, in order to avoid unwanted environmental effects which can destroy the coherence properties of the system. Because of this, deriving time-optimal control protocols is a goal of major importance. This task is usually tackled by means of quantum optimal control (QOC) theory. There, the typical problem is to derive the shape of the control field $\lambda(t)$ required in order to optimize a particular dynamical process for a quantum system described by a Hamiltonian $H(\lambda)$. For example, a typical objective in quantum control is to perform a transition from a given initial state $\Ket{\psi_0}$ to another goal state $\Ket{\psi_g}$. In some cases, the optimization can be carried out analitically \cite{bib:khaneja2001,bib:hegerfeldt2013,bib:russell2014,bib:brody2015}, but most generally its approached numerically \cite{bib:rabitz1998,bib:rabitz1998_2,bib:calarco2011}.\\

One of the weak points of the usual algorithms employed for solving QOC problems (such as Krotov or GRAPE \cite{bib:krotov1996,bib:schirmer2011}, although interesting alternatives have been proposed recently \cite{bib:bartels2013,bib:caneva2014}) is that the solution for the field $\lambda(t)$ often appears to be hardly realizable in practice \cite{bib:sundermann1999}. This originates from the fact that the value of the field at each instant acts as an independent control (i.e. there are no constraints derived from the truncation of a given basis set of functions \cite{bib:moore2012}). Of course, this allows for a much faster convergence of the optimization procedure, but the resulting control field can present non-smooth fluctuations which would require a large field bandwith to be implemented. Moreover, from a theoretical perspective, the complex shape of the field usually prevents us to understand the physical mechanisms involved in the control processes. Nevermind this feature, QOC theory has been proven to show deep connections with the fundamentals of Quantum Mechanics. Caneva \textit{et al.} \cite{bib:caneva2009} studied the performance of QOC in various systems as a function of the (fixed) evolution time $T$ that is fed to the algorithm. They found that the optimization converged succesfully only when $T$ was above certain threshold, which they identified with the quantum speed limit (QSL) time, $T_{QSL}$. The concept of QSL was introduced originally by Mandelstam and Tamm \cite{bib:mt1945}, who showed that a generalization of the usual time-energy uncertainty relation imposed bounds on the speed of evolution of a quantum system. Since then, many authors have explored the QSL in various situations \cite{bib:fleming1973,bib:bhatta1983,bib:pfeifer1993,bib:margo1998,bib:toffoli2009,bib:lloyd2013,bib:delcampo2013,bib:davidovich2013,bib:lutz2013,bib:deffner2013,bib:nos_qsl2013,bib:andersson2014}.\\

In this work we study time-optimal control processes in quantum systems which show several local two-level interactions in the form of avoided crossings (ACs) in their energy spectrum. This situation is of interest in a wide variety of quantum mechanical systems, such as molecular dipoles interacting with electric fields \cite{bib:arranz2004,bib:nos_molec2014}, ultracold atoms in optical lattices \cite{bib:tichy2013}, Rydberg atoms \cite{bib:vliegen2004} and superconducting qubits \cite{bib:dicarlo2009}. Using this particular interaction between the states of the system, we generate initial guesses for the control protocols using piecewise-constant functions derived from previous studies \cite{bib:murgida2007,bib:nos_control2013}. In a recent work, we studied the QSL time for these protocols by using QOC, and showed that the calculated QSL time is in general smaller than the sum of the optimal times for each avoided crossing \cite{bib:nos_qsl2015}. Here, we focus on the analysis of the optimal control protocols which lead to such speed-up. For that purpose, we investigate numerically the control fields that generate the time-optimal evolution and find that they can be fully characterized by just a few parameters. This allows us to propose a simple analytical dependence for the control field. Finally, for this protocol we show that the full time-evolution can be analytically solved, and the results are in full agreement with the numerical optimization.\\

This article is organized as follows. In Sec. \ref{sec:model} we present the model of a quantum system showing an avoided crossing (AC) in its energy spectrum and describe its most importante features. We then expand this minimal model to include several ACs, for which we present the actual model Hamiltonian of our interest. We also discuss the control problems that can be posed for this system, and present an intuitive solution. In Sec. \ref{sec:QOC} we present the basics of optimal control theory and describe its implementation in quantum systems. In Sec. \ref{sec:qsl_res} we briefly discuss the results obtained by implementing QOC for control processes involving several ACs, and the study the QSL time as a function of the different parameters of the system. In Sec. \ref{sec:analit} we describe in detail the optimal control fields we obtain, and discuss the physical mechanisms involved in the observed speedup. Based on this analysis, we propose a simple analytical expression for the optimal control field, and show that the associated Schr\"odinger equation can be solved analytically. Finally, Sec. \ref{sec:conclu} contains some concluding remarks. 

\section{Model, avoided crossings and control protocols} \label{sec:model}

In this section we present the models which describe the systems of our interest, which show avoided level crossings in their energy spectrum. We propose simple control protocols for achieving state transfer and discuss its time-optimality.

\subsection{A single avoided crossing}

\begin{figure}[!t]
\includegraphics[width=\linewidth]{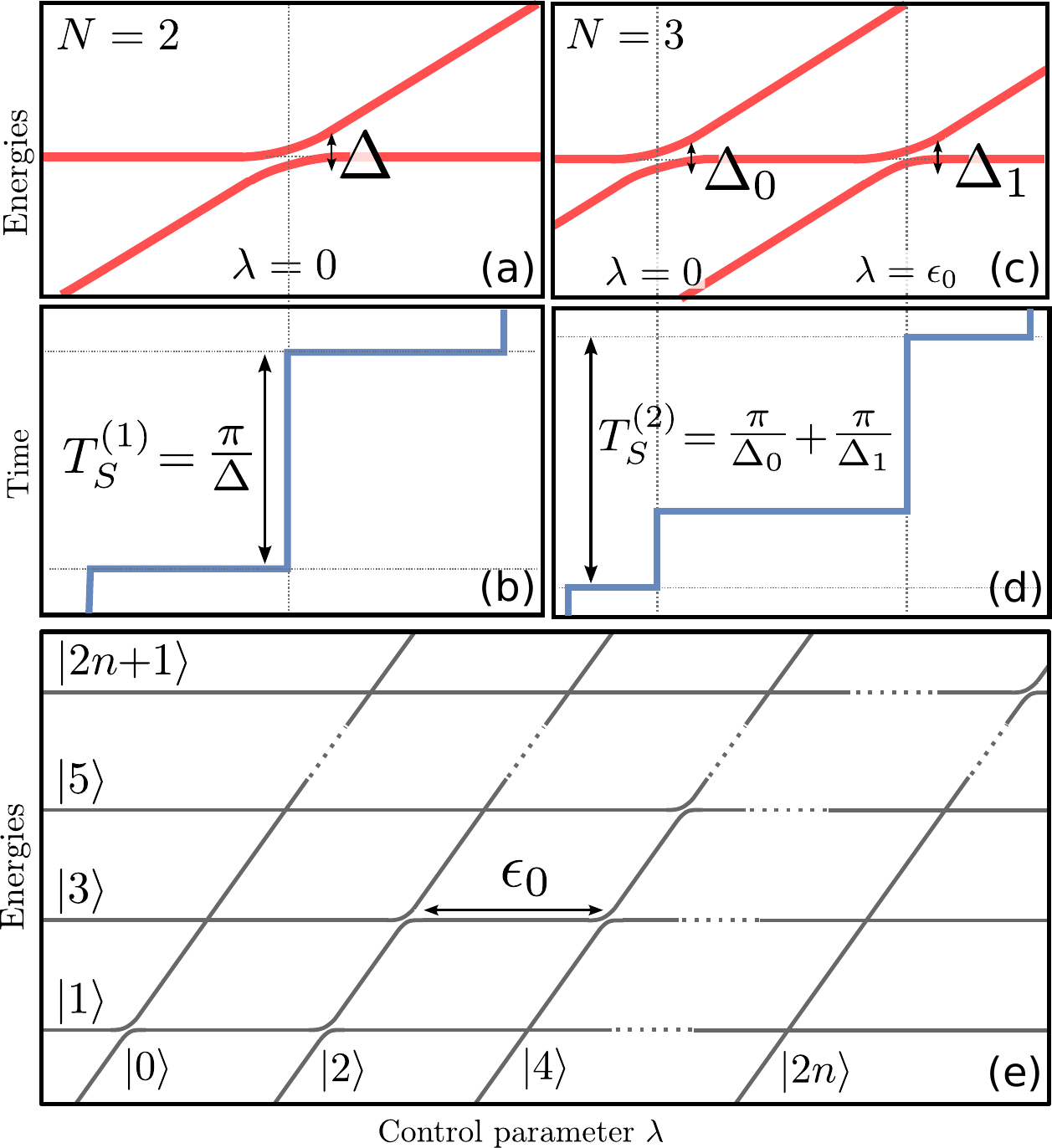}
\caption{\label{fig:fig1} (a) Energy spectrum for the two-level Hamiltonian (\ref{ec:hami2}) as a function of parameter $\lambda$. (b) Control field as a function of time for a simple realization of complete population transfer between the diabatic states. (c) and (d) same as (a) and (b) but for the three-level Hamiltonian (\ref{ec:hami3}). (e) Schematical representation of the energy spectrum of Hamiltonian $H_N(\lambda)$, c.f. Eq. (\ref{ec:hamin}), as a function of control parameter $\lambda$. In the most general setting, the spectrum shows $N-1$ avoided crossings separated by $\eps_0$, each of which generate a coupling between states $\Ket{n}$ and $\Ket{n+1}$ with $n<N-1$.  }
\end{figure}

We first consider a quantum two-level system described by the following Hamiltonian matrix

\begin{equation}
  H(\lambda)= \frac{\Delta}{2}\sigma_x+\lambda\sigma_z = \left(\begin{array}{c c}
  \lambda & \frac{\Delta}{2} \\
  \frac{\Delta}{2} & 0
  \end{array}\right),
  \label{ec:hami2}
\end{equation}

\noindent which is written in the basis $\left\{\Ket{0},\Ket{1}\right\}$. These states are usually called the diabatic states of the system, which diagonalize the Hamiltonian when the control parameter $\lambda\rightarrow\pm\infty$. In general, the eigenvalues $\{E_k\}$ ($k=0,1$) of $H$ form a hyperbolae in the $(\lambda,E)$ plane, whose vertex represents an avoided crossing (AC) with an energy gap $\Delta$. This spectrum is depicted in Fig. \ref{fig:fig1} (a). The eigenstates of $H$ as a function of $\lambda$ form the adiabatic basis $\left\{\Ket{g_\lambda},\Ket{e_\lambda}\right\}$ and have an asymptotic correspondence with their diabatic counterparts, i.e., $\Ket{g_{-\infty}}=\Ket{0}$ and $\Ket{e_{-\infty}}=\Ket{1}$ (and viceversa for $\lambda\rightarrow+\infty$).\\

The model presented above is ubiquitous in quantum mechanics as it accounts for many interesting phenomena, such as Landau-Zener transitions \cite{bib:zener1932}, Landau-Zener-Stuckelberg interferometry \cite{bib:nori2010} and quantum phase transitions \cite{bib:zurek2005}. We are interested in the control problems that can be formulated when $\Delta$ is regarded as a fixed parameter, and $\lambda$ can vary in time. A famous example is the problem of driving this system from $\Ket{g_{+\lambda_0}}$ to $\Ket{g_{-\lambda_0}}$ in the shortest possible time, for some $\lambda_0\in\mathbb{R}$. Interesting discussions about the solution to this problem, which include numerical, experimental and analytical studies can be found in the literature \cite{bib:bason2012,bib:malossi2013,bib:hegerfeldt2013,bib:nos_qsl2013,bib:hegerfeldt2014}. Here we will focus in a particular result. When $\lambda_0\rightarrow\infty$, the control problem stated above reduces to the full population transfer between $\Ket{0}$ between $\Ket{1}$. The optimal time for such process is given by

\begin{equation}
  T_S^{(1)}=\frac{\pi}{\Delta},
  \label{ec:Ts1}
\end{equation}

\noindent and can be achieved simply by setting $\lambda(t)=0$ from $t=0$ to $t=T_S^{(1)}$, given of course that $\Ket{\psi_0}=\Ket{0}$. Then, the state can be frozen in the final state for $t>T_S^{(1)}$ by applying a quench from $\lambda=0$ to some value $|\lambda|\gg\Delta$. An example of this type of control field is depicted in Fig. \ref{fig:fig1} (b).

\subsection{Multiple avoided crossings}

The two-level model described in the previous section can be extended and generalized to account for the presence of several ACs in a many-level scenario. Here we construct a model for such situation. Consider an $N$-level system with the following Hamiltonian

\begin{eqnarray}
  H_N(\lambda)&=&\sum_{n=0}^{\left[\frac{N-1}{2}\right]}\left(\lambda-n\:\eps_0\right)\KetBra{2n}{2n} \nonumber \\
  & & +\sum_{n=0}^{\left[\frac{N-2}{2}\right]}n\:\eps_0\KetBra{2n+1}{2n+1} \nonumber \\ 
  & & +\sum_{n=0}^{N-2}\frac{\Delta_n}{2}\left(\KetBra{n}{n+1}+\KetBra{n+1}{n}\right), 
  \label{ec:hamin}
\end{eqnarray}
  
\noindent where $[x]$ denotes the integer part of $x$ and $\left\{\Ket{n}\right\}$ is the basis of diabatic states. When $\Delta_n=0$ for $n=0,1,\ldots,N-2$, the Hamiltonian is diagonal in that basis, and the energy spectrum consists merely on a series of horizontal and diagonal straight branches with degeneracies at values of $\lambda_{ij}=(i+j)\eps_0$ corresponding to states $\Ket{2i}$ and $\Ket{2j+1}$. If one of the couplings is non-zero, say $\Delta_n\neq0$, the degeneracy at $\lambda_n=n\eps_0$ is lifted and an AC is generated with a minimum energy gap of $\Delta_n$. As a consequence, transitions between the states $\Ket{n}$ and $\Ket{n+1}$ become permitted. The overall shape of the energy spectrum for this model is schematically depicted in Fig. \ref{fig:fig1} (e). Note that, when all the interaction rates are non-zero, the number of ACs equals $N-1$. In a regime where $\eps_0\gg\Delta_n$ for all $n$, this model is very convenient for analyzing dynamical processes which are dictated by local two-level interactions. This can be seen as follows: if the system is initially preparred in some state $\Ket{n}$ and the control parameter $\lambda$ does not deviate much from the position of the corresponding AC (i.e. $|\lambda-\lambda_n|\ll\Delta_n$ \cite{bib:wisniacki1996}), then the dynamics of the system is effectively confined to a two-dimensional subspace, as the remaining $N-2$ levels can be adiabatically eliminated  \cite{bib:muga2010}. This is the key characteristic of our model, and we will expand on its consequences later on. \\

Evaluating Eq. (\ref{ec:hamin}) for $N=2$ we recover the two-level (one AC) Hamiltonian of Eq. (\ref{ec:hami2}). Taking the next step in complexity, the case $N=3$ renders the following Hamiltonian matrix

\begin{equation}
  H_3(\lambda)=\left(\begin{array}{c c c}
  \lambda & \frac{\Delta_0}{2} & 0 \\
  \frac{\Delta_0}{2} & 0 & \frac{\Delta_1}{2} \\
  0 & \frac{\Delta_1}{2} & \lambda-\eps_0
  \end{array}\right),
  \label{ec:hami3}
\end{equation}

\noindent which has two ACs, one at $\lambda_0=0$ and other at $\lambda_1=\eps_0$. The corresponding gaps are $\Delta_0$ and $\Delta_1$ when $\eps_0\gg\Delta_0,\Delta_1$. The energy spectrum for this case is depicted in Fig. \ref{fig:fig1} (c). This model has been widely studied in many different contexts \cite{bib:muga2010,bib:sola1999,bib:guerin2003}, as it is suitable for describing a three-level atom in a $\Lambda$ configuration. Note that, in that case, the parameters $\Delta_0$ and $\Delta_1$ correspond to detunings between the energy levels and the frequencies of two external laser fields, which are generally regarded as the control fields, while $\lambda$ and $\eps_0$ are related to the bare energy splittings. In this work this is not the case, as the off-diagonal couplings are fixed and we implement control protocols by variying solely $\lambda(t)$. \\

For this multiple AC model, we are interested in control processes which connect diabatic states of the system. Without loss of generality, we consider the initial state $\Ket{\psi_0}=\Ket{0}$ and define the process $P_K$ as the one which drives the system to the state $\Ket{K}$, with $0\leq K\leq N-1$ (generalization to a different diabatic initial state is straightforward). Our goal will be to find the control field $\lambda_K(t)$ which generates $P_K$ in a time $T$. Note that, if the ACs are sufficiently isolated, a solution exists which is independent of $N$. Based on the discussion above, an straightforward, yet powerful solution can be drawn \cite{bib:nos_control2013}. By succesively setting $\lambda(t)=\lambda_n$ constant during time intervals of length $\pi/\Delta_n$, with $0\leq n\leq K-1$ the dynamics navigates through the $K$ ACs turning them on and ensuring full population transfer one at the time. The system then evolves through the sequence $\Ket{0}\rightarrow\Ket{1}\rightarrow\ldots\rightarrow\Ket{K}$. Note that the shape of the control function is then characterized by a series of sudden changes of the value of $\lambda$, an so we name it a ``sudden switch'' field, $\lambda_K^{(S)}(t)$, which is depicted for $K=2$ in Fig. \ref{fig:fig1} (d). The total evolution time for this protocol equals

\begin{equation}
  T_S^{(K)}\equiv\sum_{n=0}^{K-1}\frac{\pi}{\Delta_n}.
  \label{ec:TsK}
\end{equation}

A number of observations are in place. First, note that the protocol proposed is not unique, since the process can also be realized by adiabatically changing $\lambda(t)$ as to navigate through the ACs. The system then also evolves sequentally between diabatic states, but much more slowly \cite{bib:murgida2007}. Also, is important to point out that we have constructed the model in Eq. (\ref{ec:hamin}) in such a way that the degeneracies between states $\Ket{n}$ and $\Ket{n+k}$ (for $k\neq1$) are exact, and cannot be lifted. For this protocol, this means that there is only one path in the energy spectrum between $\Ket{0}$ and $\Ket{K}$, which involves exactly $K$ ACs. We point out that we do not lose generality by making this assumption: if there were a shorter path between those states, it would be equivalent to a process $P_L$ with $L<K$, which is accounted for in our model.  Finally, let us remark that the total protocol time given by Eq. (\ref{ec:TsK}) is a sum which terms are of the form of Eq. (\ref{ec:Ts1}), an so we can state that the control saturates the QSL bound at each AC. In the following, we investigate wether this feature implies that the whole protocol is itself optimal or not. \\

\section{Optimal control theory} \label{sec:QOC}

Here we outline the theoretical formulation of a basic QOC problem, its solution and a feasible method for its numerical implementation. Details on this derivation can be found for example in Ref. \cite{bib:werschnik2007}.\\

Consider a quantum system described by a state $\Ket{\psi(t)}$ in a finite-dimensional Hilbert space $\mathcal{H}$ undergoing an evolution dictated by the Schr\"odinger equation (we take $\hbar=1$ from here on)

\begin{equation}
  i\frac{d}{d t}\Ket{\psi(t)}=H(t)\Ket{\psi(t)},
  \label{ec:oc_schro}
\end{equation}

\noindent satisfying $\Ket{\psi(0)}=\Ket{\psi_0}$. We supose that the Hamiltonian of the system $H(t)$ has the form

\begin{equation}
  H(t) = H_0 + \lambda(t)H_i,
\end{equation}

\noindent where $H_0$ and $H_i$ are the free (or drift) and interaction (or control) Hamiltonians, which are time-independent, and we define real-valued function $\lambda(t)$ as the control field.  The general QOC problem is formulated as follows: given $H_0$, $H_i$, an initial state $\Ket{\psi_0}$, an hermitic operator $P$ and a total evolution time $T$, we wish to find $\lambda(t)$ such that the system, initially prepared in $\Ket{\psi_0}$, evolves to a state $\Ket{\psi(T)}$, in which the expectation value of $P$ is maximal. Formally, we wish to maximize the following functional:  

\begin{equation}
  J_1\left[\psi\right]\equiv\Bra{\psi(T)}P\Ket{\psi(T)}
  \label{ec:oc_obj1}
\end{equation}

In the following we will restrict ourselves to the case in which the goal of the QOC problem is to maximize the probability of reaching a certain goal state $\Ket{\psi_g}$. For that purpose, the operator $P$ is defined as the projector $P=\KetBra{\psi_g}{\psi_g}$ and so $J_1\left[\psi\right]=\left|\BraKet{\psi(T)}{\psi_g}\right|^2$.\\

In order to correctly formulate the QOC problem, two additional conditions have to be imposed. The first one is the minimization of the quantity $\int_0^T\:\alpha(t)\lambda^2(t)dt$, where $\alpha(t)$ is a weight function. This requirement is essential in order to prevent the divergence of the total energy cost of the control process \cite{bib:rabitz1998,bib:werschnik2007}. Note that the factor $\alpha(t)$ allows for selective weighting at different times, thus allowing the induction of certain special features in the control field (e.g. a given shape, or its initial and final values \cite{bib:sundermann1999})  
As a consequence, we also wish to maximize

\begin{equation}
  J_2\left[\lambda\right]=-\int_0^T\:\alpha(t)\lambda^2(t)dt.
  \label{ec:oc_obj2}
\end{equation}

Finally, a restriction has to be imposed to the joint maximization of Eqs. (\ref{ec:oc_obj1}) and (\ref{ec:oc_obj2}), in order to guarantee that the dynamical equation (\ref{ec:oc_schro}) is satisfied at all times. For that purpose we introduce an auxiliary state $\Ket{\chi(t)}$ as a Lagrange multiplier so that we seek to maximize a third functional

\begin{equation}
  J_3\left[\chi,\psi,\lambda\right]=-2\:\mathrm{Im}\left\{\int_0^T\Bra{\chi(t)}\left(i\frac{d}{dt}-H(t)\right)\Ket{\psi(t)}dt\right\}.
  \label{ec:oc_obj3}
\end{equation}

In conclusion, bringing together expressions (\ref{ec:oc_obj1}) through (\ref{ec:oc_obj3}) we get that the QOC problem is casted as the maximization of the functional

\begin{equation}
  J\left[\chi,\psi,\lambda\right]=J_1[\psi]+J_2[\lambda]+J_3[\chi,\psi,\lambda].
  \label{ec:oc_obj4}
\end{equation}

Optimization of this functional is achieved by imposing $\delta J=0$, which renders three independent equations (one for each variable of the functional). First, solving $\delta_{\chi}J=0$ trivially gives Eq. (\ref{ec:oc_schro}), as expected from the inclusion of the Lagrange multiplier. Then, the relation $\delta_{\psi}J=0$ takes us to the following equation for the auxiliary state $\Ket{\chi}$

\begin{equation}
  i\frac{d}{d t}\Ket{\chi(t)} = H(t)\Ket{\chi(t)}\ \mathrm{and}\ \Ket{\chi(T)} = P\Ket{\psi(T)}.
  \label{ec:oc_sol1}
\end{equation}

Note that this expression is the Schr\"odinger equation for state $\Ket{\chi(t)}$, with boundary condition given by its final value, $\Ket{\chi(T)}$. Finally, by solving $\delta J_\lambda=0$  we an expression can be derived for the control field

\begin{equation}
  \lambda(t)=\frac{1}{\alpha(t)}\mathrm{Im}\left\{\Bra{\chi(t)}H_i\Ket{\psi(t)}\right\}.
  \label{ec:oc_sol2}
\end{equation}


The problem of obtaining a set $\left\{\Ket{\psi(t)},\Ket{\chi(t)},\lambda(t)\right\}$ that simultaneously solve Eq. (\ref{ec:oc_schro}) together with Eqs. (\ref{ec:oc_sol1}) and (\ref{ec:oc_sol2}) is, of course, impossible to tackle analytically. Instead, an iterative algorithm has to be implemented. Here we briefly describe a widely used method \cite{bib:somloi1993,bib:montangero2007}, due originally to Krotov \cite{bib:krotov1996}: (i) the procedure starts by choosing an initial guess $\lambda_0(t)$ for the control field; (ii) using that field, the initial state $\Ket{\psi_0}$ is evolved according to Eq. (\ref{ec:oc_schro}) from $t=0$ to $t=T$; (iii) the boundary condition for $\Ket{\chi(t)}$ is set by projecting $\Ket{\chi(T)}=P\Ket{\psi(T)}$, and the state is evolved backwards also following Eq. (\ref{ec:oc_schro}), from $t=T$ to $t=0$; (iv) the state $\Ket{\psi_0}$ is now propagated forward again, but the field is updated following the rule $\lambda(t)\rightarrow\lambda(t)+\frac{1}{\alpha(t)}\mathrm{Im}\left\{\Bra{\chi(t)}H_i\Ket{\psi(t)}\right\}$ at each instant; (v) steps (iii) and (iv) are repeated $\mathcal{N}$ times until a certain threshold is reached for the value of the cost functional $J_1$.\\


\section{QSL in a system with multiple ACs} \label{sec:qsl_res}

In this section we numerically investigate the QSL time for the control processes described in the previous section. For that purpose we use optimal control techniques, inspired by the basic idea introduced by Caneva et al. \cite{bib:caneva2009} that the optimization performance is limited by the maximum speed allowed by quantum evolution. The basic procedure is as follows. First, we fix the state dimension $N$ and choose a control process $P_K$ for the model described in Section \ref{sec:model}. Then, we run the optimization algorithm in order to find the control field $\lambda_K(t)$ which generates the desired process, for different values of the total evolution time $T$. In each run, this procedure takes as an input the value of $T$ and an initial guess for the field $\lambda_K^{(0)}(t)$. In order to choose these inputs, we take advantage of the physical features of the model discussed in the previous section. The values of $T$ were taken from an interval centered around $T_S^{(K)}$, cf. Eq. (\ref{ec:TsK}). Note that, if the ACs are well isolated, we are certain that the sudden switch field generates the desired process when $T=T_S^{(K-1)}$. Similarly, the initial guess for the control function were chosen to be close to the sudden switch field. Actually, we used

\begin{equation}
  \lambda_K^{(0)}(t) = a(t)\lambda_K^{(S)}(b\:t)+c(t)
  \label{ec:initialguess}
\end{equation}

\noindent where $b$ is a parameter which shrinks or expands the shape of the function to fit the total evolution time (i.e. $b=1$ when $T=T_S^{(K)}$), while $a(t)$ is a function which smooths the discontinuities of $\lambda_K^{(S)}$ and $c(t)$ is a small linear correction. The latter functions are introduced in order to force the algorithm to take a minimum number of steps (of the order of 100) before the required convergence is achieved. \\

\begin{figure}[!t]
\includegraphics[width=\linewidth]{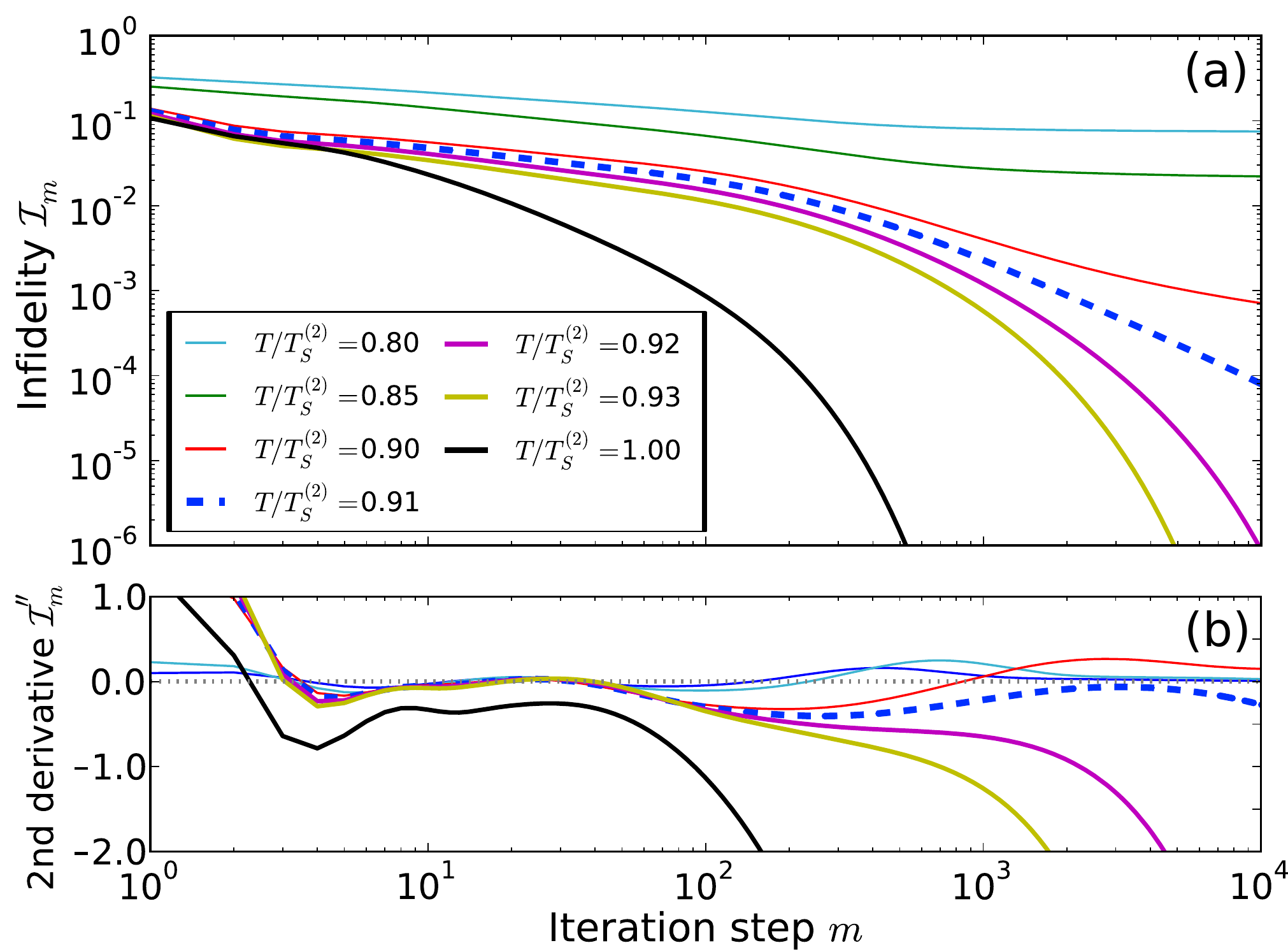}
\caption{\label{fig:fig2} (color online) A typical procedure for determining the QSL time $T_{QSl}$ for a particular control process. This example corresponds to a $P_{K=2}$ process in a Hilbert space of dimension $N=3$, cf. Eq. (\ref{ec:hami3}), where $\eps_0=10\Delta_0$ and $\Delta_1=\Delta_0$. (a) Infidelity $\mathcal{I}_m$ as a function of the step number $m$ for each optimization. (b) Second derivative $\mathcal{I}''_m$ of the curves in (a). The dotted line corresponds to the minimum value of $T$ which asymptotically renders $\mathcal{I}''_m<0$ and so is identified as $T_{QSL}$  within error margin. Thin full lines correspond to $T<T_{QSL}$ while, thick full lines to $T>T_{QSL}$.}
\end{figure}

Each run of the algorithm finishes after a fixed number of steps, or when the process is sufficiently converged. As discussed in Section \ref{sec:QOC}, this is determined by evaluating the value of the infidelity at each step $m$, which is defined as

\begin{equation}
  \mathcal{I}_m\equiv1-\left|\BraKet{\psi_g}{\psi^{(m)}(T)}\right|^2=1-J_1[\psi^{(m)}(T)],
  \label{ec:infid}
\end{equation}

\noindent where $J_1$ is the functional of Eq. (\ref{ec:oc_obj1}), and $\Ket{\psi^{(m)}(t)}$ is the state of the system obtained at step $m$ of the algorithm. The function $\mathcal{I}_m$ decreases monotonically as $m$ increases, but its shape and asymptotic behaviour depends critically on the input parameters. In Fig. \ref{fig:fig2} (a) we plot this function for a particular case, as an example. We argue, as in Ref. \cite{bib:caneva2009} that the infidelity cannot decrease indefinitely if $T$ is smaller than the QSL time. In that case, $\mathcal{I}_m$ should look asymptotically flat. We use this feature to obtain the estimator of the QSL time $T_{QSL}^{(K)}$. Formally, for each value of $T$ we look at the second derivative of $\mathcal{I}_m$ (with respect to $m$), see Fig. \ref{fig:fig2} (b), and analyze its sign. Then, the minimum value of $T$ which gives $\mathcal{I}''(k)<0$ asymptotically, is chosen as the QSL time.\\

\begin{figure}[t]
\includegraphics[width=\linewidth]{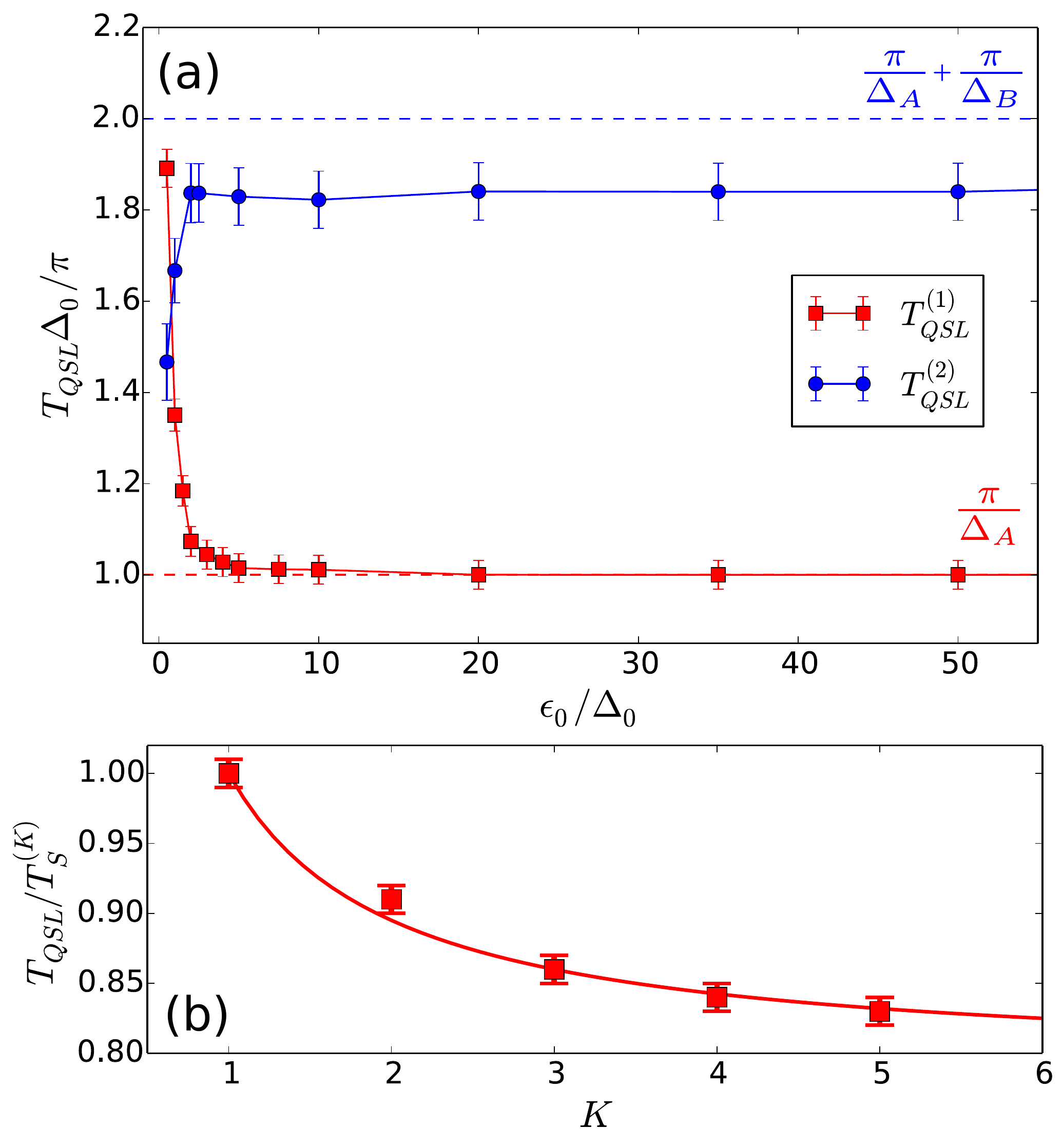}
\caption{\label{fig:fig3} (color online) (a) QSL time calculated from the optimal control procedure (see text for details) as a function of $\eps_0/\Delta_0$, for processes $P_1$ (crossing one AC) and $P_2$ (crossing two ACs). Dashed lines correspond to expression (\ref{ec:TsK}), i.e. the time required by the sudden-switch protocol in each case, $T_S^{(1)}$ and $T_S^{(2)}$. (b) Ratio between the calculated QSL time $T_{QSL}$ for processes $P_K$ and the corresponding sudden-switch protocol evolution time $T_S^{(K)}$ as a function of $K$. The dashed line show the $K^{-1}$ scaling of the data points. For all cases, the distance between the ACs was set to $\eps_0=10\Delta$, and $\Delta_n=1$ for all $n=0,1,\ldots,K-1$.  }
\end{figure}

We now turn our focus to the model of Eq. (\ref{ec:hami3}) which presents two ACs. As discussed in the previous Section, this is the next step in complexity following the analytically solvable two-level system. We begin by considering the QSL time for process $P_1$, for which the system starts in state $\Ket{0}$ and evolves to $\Ket{1}$, in the minimum possible time. Note that this process involves just one AC, as seen from the sudden-switch protocol introduced in the previous section. In Fig. \ref{fig:fig3}, we plot the calculated QSL time $T_{QSL}^{(1)}$ for this case as a function of $\eps_0$, the parameter which measures the distance between the ACs in the energy spectrum (see Fig. \ref{fig:fig1}), for fixed values of interaction parameters $\Delta_0,\Delta_1$. There, it can be seen that $T_{QSL}^{(1)}$ is larger than $T_S^{(1)}=\pi/\Delta_0$ for small values of $\eps_0$. This is reasonable in this regime, since the ACs interact considerably, which leads to significant variations of the interaction rates (see Ref. \cite{bib:muga2010} for more details). Away from that regime, $T_{QSL}^{(1)}$ converges to $T_S^{(1)}$, which is the well-known result for the two-level system. This is a sound result, since only the states $\Ket{0}$ and $\Ket{1}$ are involved in the process. However, it is interesting to point out that this behaviour allows us to quantitatively define the regime in which the ACs are well isolated. In the case shown in the figure, for which $\Delta_0=\Delta_1$, this is achieved for $\eps_0/\Delta_A\gtrsim5$.\\

Next, we discuss control process $P_2$, which involves both ACs. Following the same procedure as for the previous case, we get the results of Fig. \ref{fig:fig3} (a). There, it can be seen that the estimated QSL time $T_{QSL}^{(2)}$ is smaller than the sudden switch evolution time $T_S^{(2)}$. Remarkably, this result holds in all cases, even for large $\eps_0$. The difference between $T_{QSL}^{(2)}$ and our prediction is larger for small $\eps_0$, and decreases as the ACs are brought apart. However, for $\eps_0/\Delta_0$ as large as $100$, the difference is still larger than $7\%$. This striking behaviour indicates that the QOC optimization can generate successful (i.e., with arbitrary fidelity) control processes which are significantly shorter in time than the double sudden-switch, a process wich is time-optimal at each AC, as discussed above. We point out that this behaviour persists even when the relative magnitude of the gap sizes is modified \cite{bib:nos_qsl2015}. We will analyze the physical mechanisms that cause this speed-up in the next section. \\

Finally, we address the results obtained for the QSL time for control processes involving more than two ACs, i.e. $P_K$ with $K>2$. Applying the same procedure outlined in the previous paragraphs, we obtained $T_{QSL}$ for various values of the number of avoided crossings $K$ involved in the process. In Fig. \ref{fig:fig3} (b) we plot the ratio between $T_{QSL}$ and $T_S^{(K)}$ as a function of $K$. There, it can be seen that the optimal evolution time (measured with respect to the corresponding sudden switch protocol evolution time) decreases as the number of ACs involved increases. This means that, as more ACs get involved in the evolution, the connection between diabatic states can be performed faster. However, the improvement reaches a saturation point for large values of $K$.

\section{Optimal control fields} \label{sec:analit}

\subsection{Numerical analysis of the optimized fields}

\begin{figure}[t]
\includegraphics[width=\linewidth]{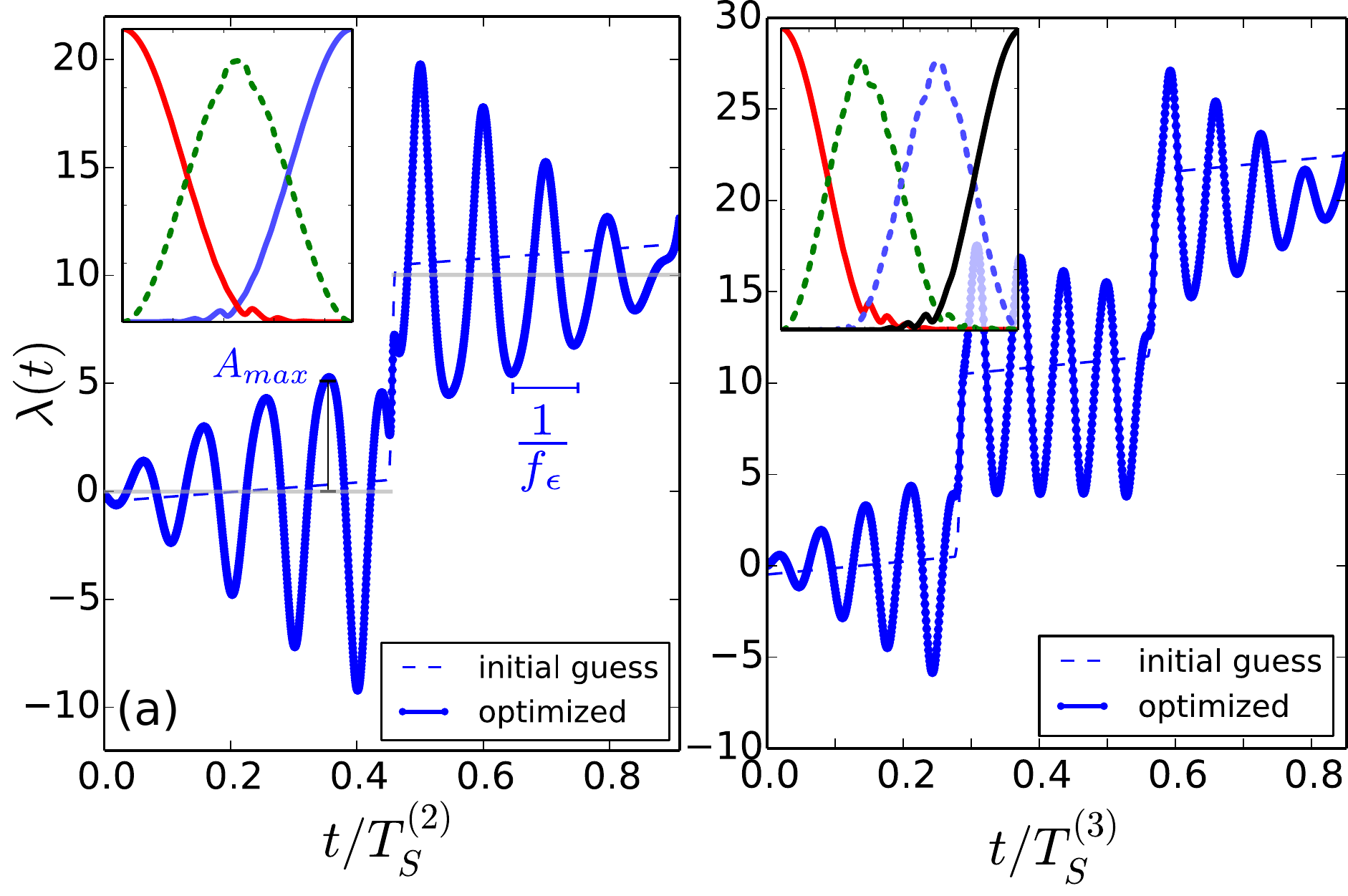}
\caption{\label{fig:fig4} (color online) (a) Initial and optimized control fields $\lambda(t)$ for process $P_2$, using $\Delta_1/\Delta_0=1$ and $\eps_0/\Delta_0=10$. Inset show the time evolution of the populations for each one of the diabatic states $\Ket{k}$ ($k=0,1,2$), given by the optimized field. Evolution time is set at $T=T_{QSL}^{(2)}\simeq0.91 T_S^{(2)}$. (b) same as (a) for process $P_3$, with same parameter values and $\Delta_2/\Delta_0=1$ as well. Evolution time is set at $T=T_{QSL}^{(3)}\simeq0.85 T_S^{(3)}$.}
\end{figure}

\begin{figure}[t]
\includegraphics[width=\linewidth]{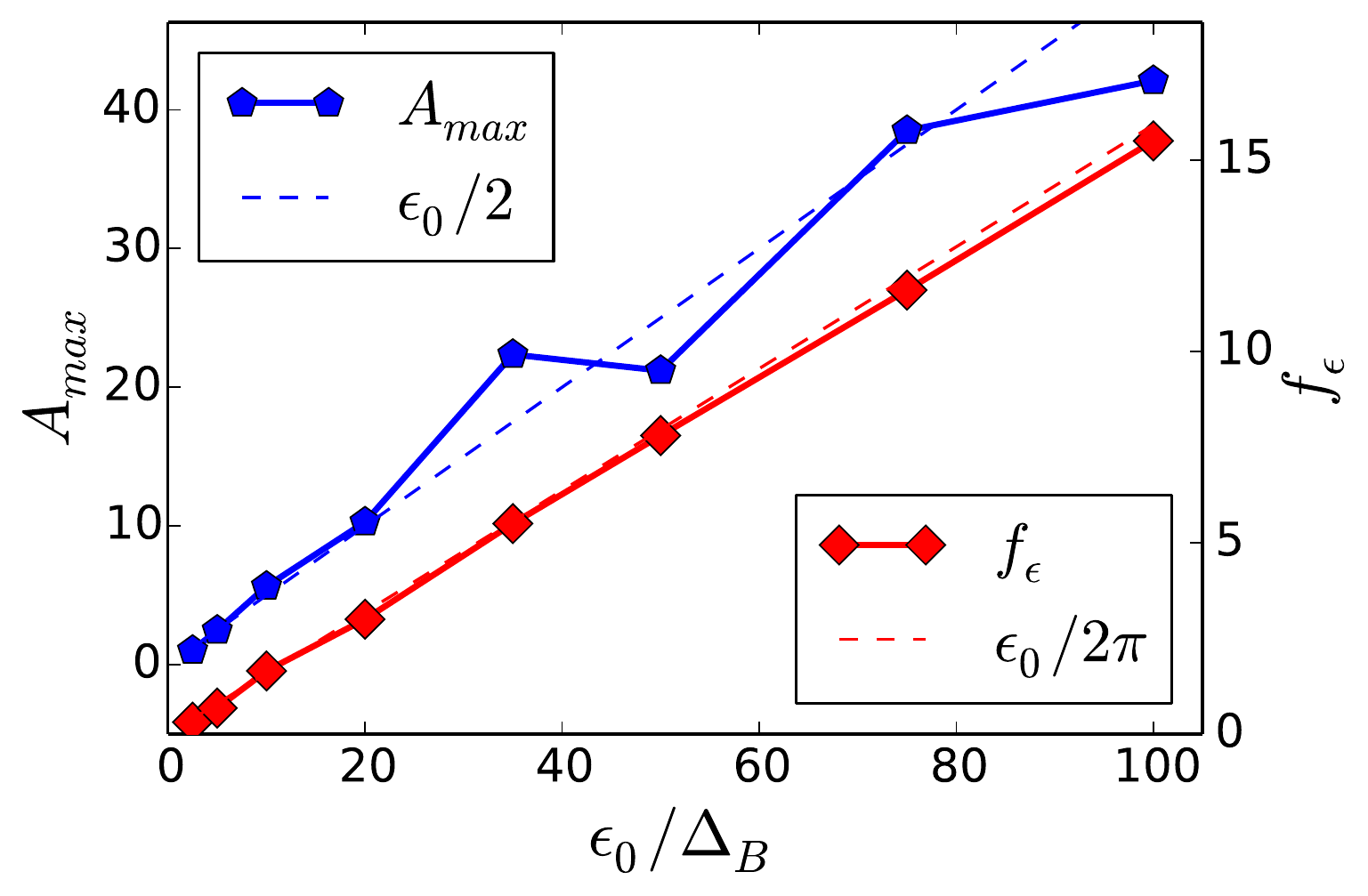}
\caption{\label{fig:fig5} (color online) Frequency $f_\eps$ (right axis) and maximum oscillation amplitude $A_{max}$ (left axis) of the optimal field for process $P_2$ (see Fig. \ref{fig:fig4}) as a function of $\eps_0$. Dashed lines indicate lineal dependences of both quantities with $\eps_0$.}
\end{figure}

We now turn to analyze the shape of the control fields derived via the optimization procedure outlined in the previous section. We will focus on the optimal fields obtained for $T=T_{QSL}$, but for larger evolution times its description is similar. In Fig. \ref{fig:fig4} (a) and (b) we plot the optimized field $\lambda(t)$ together with the evolution of the populations for two particular cases, with $K=2$ and $K=3$. At first sight, it can be seen that the field shows oscillations wich are mounted on a step-like function. The latter feature is preserved from the sudden-switch field, wich we used as an initial guess for the optimization. Fourier transform of the driving signal reveals that there is only one dominant frequency $f_\eps$, which together with the maximum amplitude $A_{max}$, characterizes the overall shape of the field. Remarkably, this behaviour is common to all high-order control processes studied in our model, even for $K>3$. In order to quantitatively analyze the driving field, we studied the dependance of $f_\eps$ and $A_{max}$ as a function of the distance $\eps_0$ between the ACs. Results are shown in Fig. \ref{fig:fig5} for $K=2$, where the linear dependence of both quantities with $\eps_0$ is clear and can even be regarded as exact for the frequency, for which we can write $f_\eps=\eps_0/2\pi$.\\

The regular behaviour shown by the numerically optimized control field has some interesting consequences. First, note that as $\eps_0$ increases and the avoided crossings get further apart, the driving field will require a bigger intensity and a larger bandwith in order to be implemented. In practice, at some point this requirement will no longer be fulfilled, and most likely the QSL time will tend to $T_S^{(K)}$ for all practical purposes. This is indeed reasonable, since technical limitations would then imply that the ACs are effectively isolated, with no possible coupling between them. However, from a theoretical standpoint, this is a much different scenario than the one usually obtained in QOC optimizaton, where the broad bandwidth requirements originates from the highly irregular features of the optimized field. In our case the control function $\lambda(t)$ can be readily described by a few parameters. We associate this remarkable feature with the special characteristics of our model, which shows localized two-level interactions in a many-level spectrum, a scenario which is common in many different physical setups, as previously mentioned. \\

\subsection{Analytical approximate solution for the time-dependent problem}

The shape of the field also gives us interesting insight about the physical mechanisms involved in the observed enhancement of the QSL time for these control processes \cite{bib:nos_qsl2015}. Remarkably, we found that an anallytical approximation for the time-dependent evolution can be drawn inspired from the results of the optimization process. We will show this solution in the following for the case $K=2$, although the idea can extended higher order processes. Recall the Hamiltonian $H_3(\lambda)$ from Eq. (\ref{ec:hami3}), which can be written as the sum of its non-diagonal and diagonal parts

\begin{equation}
  H_3(\lambda)=H_{ND}+H_D(\lambda),
  \label{ec:hami3nd}
\end{equation} 

\noindent in such a way that $H_{ND}$ depends on the coupling parameters $\Delta_0$ and $\Delta _1$ while the dependence on the control parameter is concentrated in $H_D(\lambda)$. We propose the following expression for the driving field

\begin{equation}
  \lambda(t)=\left\{\begin{array}{l l l}
  \lambda_A\cose{\omega t+\phi} & , & 0\leq t<t_m \\
  \eps_0 + \lambda_A\cose{\omega (t-t_m)+\tilde{\phi}} & , & t_m\leq t \leq T 
  \end{array}\right..
  \label{ec:campo}
\end{equation}

This field has the form of a step-wise constant function with oscillations of angular frequency $\omega$ mounted on each step (note that, from the previous analysis, we can infere that $\omega=\eps_0$). The field then oscillates around a fixed value at each step, corresponding to the localization of the two ACs: a $t=0$ it begins at $\lambda=\lambda_0=0$, and then turns to $\lambda=\lambda_1=\eps_0$ at some $t=t_m$. The overall shape of $\lambda(t)$ then emulates the optimized field seen in Fig. \ref{fig:fig4} (a), with the difference that we use a constant amplitude $\lambda_A$ for the oscillating term, for convenience.\\

Let us first consider the dynamics from $t=0$ to $t=t_m$. We propose that the total evolution operator for this evolution can be factorized as

\begin{equation}
  U_0(t,0)=U_0^{(A)}(t)U_0^{(B)}(t),
  \label{ec:evol}
\end{equation}

\noindent where $U_0{(A)}(t)=\expo{-i\int_0^{t_m}H_D(t')dt'}$ is diagonal in the diabatic basis and the superscript emphasizes the fact that we are working on the AC located at $\lambda=\lambda_0=0$. The problem is then to find the unitary operator $U^{(B)}(t)$, which satisfies the Schr\"odinger equation in the interaction picture $i\dot{U}^{(B)}(t)=\tilde{H}_{ND}(t) U^{(B)}(t)$, with $\tilde{H}_{ND}\equiv U_0^{(A)\dagger}H_{ND}U_0^{(A)}$ being the corresponding transformed Hamiltonian, which takes the form

\begin{equation}
  H_{ND}'(t)=\frac{\ex{-i\lambda_i}}{2}\left(\begin{array}{c c c}
  0 & \ex{2i\lambda_i}\Delta_0 & 0 \\
  \Delta_0 & 0 & \ex{i\eps_0t}\Delta_1 \\
  0 & \ex{2i(\lambda_i-\eps_0t)}\Delta_1 & 0
  \end{array}\right),
  \label{ec:hndmono0}
\end{equation}

\noindent where we have defined $\lambda_i\equiv\lambda_i(t)=\int_0^{t_m}\lambda(t')dt'=\frac{\lambda_A}{\omega}\seno{\omega t+\phi}-\phi_0$ and $\phi_0=\frac{\lambda_A}{\omega}\seno{\phi}$. The unitary evolution problem is then casted in terms of this time-dependent Hamiltonian. The key to consider here is that the exponentials that appear in the previous expression can be written in Fourier series using the identity 

\begin{equation}
  \ex{i z \sin\gamma}=\sum_{n=-\infty}^{n=\infty}J_n\left(z\right)\ex{in\gamma},
\end{equation}

\noindent where $J_n(z)$ simbolyzes the Bessel $J$-function of order $n$. The results we obtained from the QOC procedure indicates us that the frequency of the driving at each step $\omega$ is much larger than $\Delta_0,\Delta_1$. Then, most of the terms in Eq. (\ref{ec:hndmono0}) oscillate very quickly and can thus be neglected. This kind of rotating wave approximation is typically invoked when analyzing high-frequency modulation of periodic potentials \cite{bib:nori2007}, for example in a cold atoms setup \cite{bib:creffield2014}. Formally, we approximate

\begin{eqnarray}
  \ex{-i(\lambda_i(t)-\eps_0t)}&=&\ex{i\phi_0}\sum_n J_n\left(\frac{\lambda_A}{\omega}\right)\ex{-i(n\omega-\eps_0)-in\phi} \\ \nonumber 
  & = & \ex{i(\phi_0-\phi)}J_1\left(\frac{\lambda_A}{\omega}\right)\Delta_1,
\end{eqnarray}

\noindent where we used the argument of the previous paragraph to identify the term $n=1$ as the resonant one and set $\omega=\eps_0$, which was expected from the numerical analysis of the optimal fields. Its straightforward to calculate the rest of the elements of Hamiltonian of Eq. (\ref{ec:hndmono0}), which can be approximated by a time-independent expression

\begin{equation}
 H_{ND}'=\frac{1}{2}\left(\begin{array}{c c c}
  0 & \ex{-i\phi_0}\Delta_0' & 0 \\
  \ex{i\phi_0}\Delta_0' & 0 & \ex{i(\phi_0-\phi)}\Delta_1' \\
  0 & \ex{-i(\phi_0-\phi)}\Delta_1' & 0
  \end{array}\right),
  \label{ec:hndmono0_bis}
\end{equation}

\noindent where we have introduced the renormalized interaction rates

\begin{eqnarray}
  \Delta_0' & \equiv & J_0\left(\frac{\lambda_A}{\eps_0}\right)\Delta_0 \nonumber \\
  \Delta_1' & \equiv & J_1\left(\frac{\lambda_A}{\eps_0}\right)\Delta_1 \label{ec:renorm}
\end{eqnarray}

Then, the evolution of the system for $0\leq t <t_m$ is completely determined by the evolution operator in Eq. (\ref{ec:evol}) where $U_0^{(B)}(t)=\expo{-iH_{ND}'t}$. Note that this factor introduces the couplings between the diabatic states which generate the time evolution of the operators. The role of the driving in this process is clear. In the absence of the oscillatory field, i.e. $\lambda_A=0$, Eq. (\ref{ec:renorm}) gives $\Delta_0'=\Delta_0$ and $\Delta_1'=0$, and so only states $\Ket{0}$ and $\Ket{1}$ can be connected in this evolution. This is exactly what we expected from the sudden-switch protocol and the adiabatic elimination procedure discussed in Section \ref{sec:model}. When the oscillatory field is turned on, $\Delta_0$ decreases and $\Delta_1$ takes a non-zero value, thus coupling weakly states $\Ket{1}$ and $\Ket{2}$. This generates an evolution where the goal state of the protocol $\Ket{2}$ can draw a portion of the population of the other levels even when the dynamics is mainly dictated by the first AC. Thanks to this feature, the evolution towards the goal state is accelerated, thus providing the overall enhancement of the QSL time shown in the previous Section. Note that Eq. (\ref{ec:renorm}) formalizes the fact that the amplitude $\lambda_A$ cannot be neglected with respect to $\eps_0$. Moreover, setting $\lambda_A/\eps_0$ to a constant value is consistent with the analysis shown in Fig. \ref{fig:fig4} (c), where both quantites showed a linear correlation.\\

\begin{figure}[t]
\includegraphics[width=\linewidth]{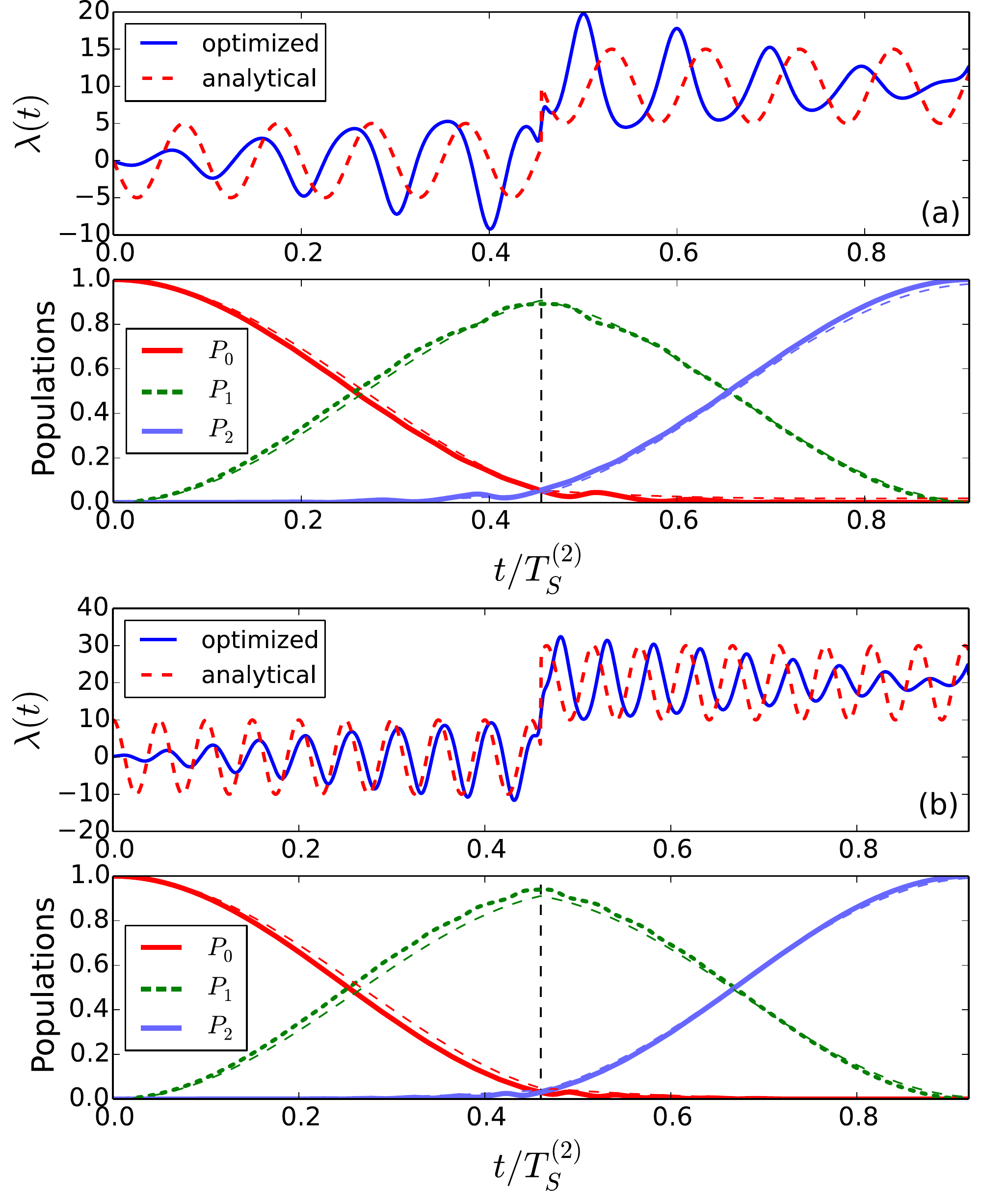}
\caption{\label{fig:fig6}(color online) Top: Optimized and analytical proposal of Eq. (\ref{ec:campo}) for the control field as a function of time. Bottom: time-evolution of the populations for each of the diabatic states. Thin dashed lines corresponds to the analytical solution of Eq. (\ref{ec:utotal}). (a) Case $\eps_0=10\:\Delta_0$. (b) Case $\eps_0=20\:\Delta_0$.}
\end{figure}

An analogous procedure can be done for the evolution between $t=t_m$ and $t=T$, in such a way that we can finally write for the whole evolution as

\begin{equation}
  U(t) = \left\{\begin{array}{c c c}
  U_0^{(A)}(t)U_0^{(B)}(t) &,& 0\leq t<t_m \\
  U_1^{(A)}(t)U_1^{(B)}(t)\times & & \\
  U_0^{(A)}(t_m)U_0^{(B)}(t_m) &,& t_m\leq t\leq T
  
  \end{array}\right.,
  \label{ec:utotal}
\end{equation}

\noindent where we have defined $U_1^{(A)}(t)=\expo{-i\int_{t_m}^{T}H_D(t')dt'}$ and $U_1^{(B)}(t)=\expo{-iH_{ND}'' (t-t_m)}$ with an effective time-independent Hamiltonian given by

\begin{equation}
 H_{ND}'=\frac{1}{2}\left(\begin{array}{c c c}
  0 & \ex{-i(\tilde{\phi_0}-\tilde{\phi})}\Delta_0'' & 0 \\
  \ex{i(\tilde{\phi_0}-\tilde{\phi})}\Delta_0'' & 0 & \ex{i\tilde{\phi_0}}\Delta_1'' \\
  0 & \ex{-i\tilde{\phi_0}}\Delta_1'' & 0
  \end{array}\right).
  \label{ec:hndmono1}
\end{equation}

The renormalized interaction rates are now interchanged with respect to the previous case, 

\begin{eqnarray}
  \Delta_0'' & \equiv & J_1\left(\frac{\lambda_A}{\eps_0}\right)\Delta_0 \nonumber \\
  \Delta_1'' & \equiv & J_0\left(\frac{\lambda_A}{\eps_0}\right)\Delta_1, \label{ec:renorm2}
\end{eqnarray}

\noindent which is natural since in the second step the dominant interaction is due to the AC between states $\Ket{1}$ and $\Ket{2}$. In Fig. \ref{fig:fig6} we show the time evolution of the populations for different cases, as predicted by the analytical formula (\ref{ec:utotal}). There, it can be seen how this expressions approximates very well the optimized evolution, even though the driving fields are not exactly equal. This behaviour allows us to assert that the high-frequency oscillations of the driving field at each AC, switches on the adjacent ACs allowing for the population of other energy levels and thus providing the mechanism for the overall speed-up of the control processes. Note that the solution we provide here is based on the process $P_{K=2}$, being the next step in complexity of the two-level case, where this novel effects are, of course, absent. We believe that a similar procedure could be applied to find analytical expressions for higher-order processes.\\

\subsection{Initial guess and performance of the optimization}

\begin{figure}[t]
\includegraphics[width=\linewidth]{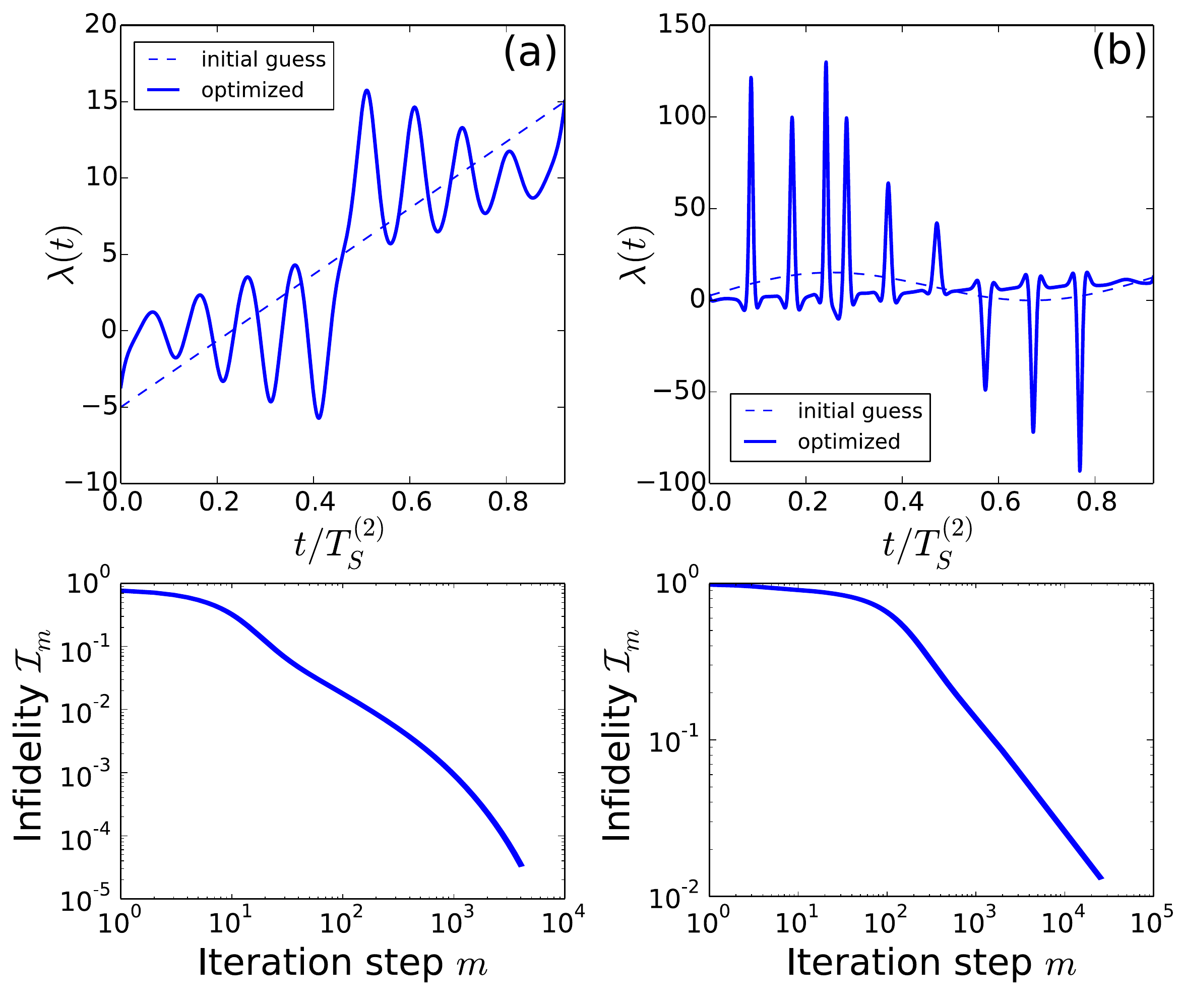}
\caption{\label{fig:fig7} Top: Optimized control field as a function of time. Bottom: Infidelity $\mathcal{I}_m$ as a function of the step number $m$ for each optimization. (a) Initial guess is a linear function. (b) Initial guess is a sinusoidal function.}
\end{figure}

We recall that the previous analytical discussion was motivated by the fact that the control field obtained from QOC had a simple shape. This, in turn, related with the fact that the optimized field preserved certain features of the initial guess $\lambda_0(t)$ we employed, for example, the step-wise structure. We will now briefly discuss the role of the initial guess in the optimization. Note that the results we showed in Section \ref{sec:qsl_res} are independent of the initial guess we propose for the field. However, this election changes the overall performance of the optimization. In Fig. \ref{fig:fig7} we show the infidelity and optimized field obtained for two choices of $\lambda_0(t)$ different from the one shown in Fig. \ref{fig:fig4}. In one of the cases, Fig. \ref{fig:fig7} (a), the field has a linear dependence and connects the positions of the ACs. The corresponding optimized field develops fast oscillations and overall looks very similar to the one in Fig. \ref{fig:fig4}. Moreover, the frequency of the oscillations $f_\eps$ is the same for both cases, and the total number of iterations required for the convergence of the algorithm is also of the same order (around 4000). On the other hand, in Fig. \ref{fig:fig7} (b) we used as an initial guess a sinusoidal field, with initial and final value at $\lambda=\lambda_0=0$, and for which we deliberately change the parity with respect to the other cases. In that case, the number of iterations required by the optimization to converge raises by a factor of 10. Moreover, the optimized fields has a very irregular shape, showing peaks of very large amplitude (50 times bigger than the other cases). \\

As we mentioned previously, the fact that the QSL time for control processes can be drawn from the optimization procedure enforces the power of QOC as a tool in this context. The results we show here also tells us that the performance of QOC, and its ability to give us information about the physical mechanisms involves in a control processes can be enhanced by properly providing the optimization with a good initial guess. In this case we have done so by analyzing the characteristics of the system, and more precisely by studying the structure of the energy spectrum.\\

\section{Final remarks} \label{sec:conclu}

In this work we studied time-optimal quantum control in systems with multiple avoided crossings in their energy spectrum. Based on previous works \cite{bib:murgida2007,bib:nos_control2013}, we are able to ellaborate protocols which generated the desired control processes. We used these protocols as initial guesses for an optimization procedure which leads us to several results. Following recent work \cite{bib:nos_qsl2015}, we discussed how the quantum speed limit time for these systems as a function of the separation between the AC is enhanced with respect to the result derived from the two-level approximation. We observed that when the dimension of the system is increased (and with it, the number of ACs involved in the dynamical process) the speed-up becomes more pronounced. Having obtained the control protocols at the QSL, we numerically analyzed the shape of the fields derived from the optimization. We found that they showed a very regular behaviour, characterized by the presence of single-frequency oscillations mounted on a step-wise function. Based on these results, we were able to construct a model for the control problem which we solved analytically. This model also allows us to qualitatively explain that the main feature behind the optimization was the collective dynamics of multiple avoided-crossings. Finally, we studied how the outcome and performance of the optimization were modified when varying the initial guess for the control protocol. We found that using different initial guesses can lead to very different shapes of the optimized control field. This behaviour indicates that a preliminary analysis of the system spectrum, as done here, can act as pre-optimization method as it lead us to a good choice of the initial guess.

\begin{acknowledgments}

We acknowledge support from CONICET, UBACyT, and ANPCyT (Argentina).

\end{acknowledgments}


\begin{thebibliography}{60}
\small{

\bibitem{bib:shapiro2011}  M. Shapiro and P. Brumer, {\it Quantum Control of Molecular Processes}, (Wiley-VCH, Berlin, 2011).

\bibitem{bib:wiseman2009} H.M. Wiseman and G.J. Milburn, {\it Quantum Measurement and Control}, (Cambridge, 2009).

\bibitem{bib:dalessandro2008} D. D'Alessandro, \textit{Introduction to Quantum Control and Dynamics}. (Chapman \& Hall/CRC, 2008).

\bibitem{bib:meshulach1998} D. Meshulach and Y. Silberberg, Nature (London) {\bf 396},  239 (1998).

\bibitem{bib:press2008} D. Press, T. Ladd, B. Zhang and Y. Yamamoto, Nature (London) {\bf 456}, 218 (2008).

\bibitem{bib:khaneja2001} N. Khaneja, R. Brockett and S.J. Glaser, Phys. Rev. A \textbf{63}, 032308 (2001)

\bibitem{bib:hegerfeldt2013} G.C. Hegerfeldt, Phys. Rev. Lett. {\bf 111}, 260501 (2013).
  
\bibitem{bib:russell2014} B. Russell and S. Stepney, Phys. Rev. A {\bf 90}, 012303 (2014).

\bibitem{bib:brody2015} D.C Brody, G.W. Gibbons and D.M. Meier, D.M., New J. Phys. {\bf 17}, 033048 (2015).
  
\bibitem{bib:rabitz1998} W. Zhu, J. Botina and H. Rabitz, J. Chem. Phys. {\bf 108}, 1953 (1998).
  
\bibitem{bib:rabitz1998_2} W. Zhu and H. Rabitz, J. Chem. Phys. \textbf{109}, 385 (1998).

\bibitem{bib:calarco2011} P. Doria, T. Calarco and S. Montangero, Phys. Rev. Lett. {\bf 106} 190501, (2011).

\bibitem{bib:krotov1996} V. F. Krotov, \textit{Global Methods in Optimal Control Theory} (Marcel Dekker, New York, 1996)

\bibitem{bib:schirmer2011} S.G. Schirmer and P. de Fouquieres, New J. Phys. \textbf{13} 073029, (2011).

\bibitem{bib:bartels2013} B. Bartels and F. Mintert, Phys. Rev. A \textbf{88} 052315, (2013).

\bibitem{bib:caneva2014} T. Caneva, T. Calarco and S. Montangero, Phys. Rev. A \textbf{84} 022326, (2014).

\bibitem{bib:sundermann1999} K. Sundermann and R. de Vivie-Riedle, J. Chem. Phys. \textbf{110} 1896, (1999).

\bibitem{bib:moore2012} K. Moore and H. Rabitz, J. Chem. Phys. \textbf{137}, 134113 (2012).

\bibitem{bib:caneva2009} T. Caneva \textit{et al.}, Phys. Rev. Let. {\bf 103}, 240501 (2009).

\bibitem{bib:mt1945} L. Mandelstam and I. Tamm, J. Phys. USSR {\bf 9}, 249 (1945).

\bibitem{bib:fleming1973} G.N. Fleming, Nuov. Cim., {\bf 16 A}, 232, (1973).
  
\bibitem{bib:bhatta1983} K. Bhattacharyya, J. Phys. A {\bf 16} 2993, (1983).

\bibitem{bib:pfeifer1993} P. Pfeifer, Phys. Rev. Lett. {\bf 70} 33653368, (1998).
  
\bibitem{bib:margo1998} N. Margolus and L.B. Levitin, Physica D {\bf 120} 188-195, (1998).
  
\bibitem{bib:toffoli2009} L.B. Levitin and T. Toffoli, Phys. Rev. Lett. {\bf 103} 160502, (2009).

\bibitem{bib:lloyd2013} V. Giovannetti and S. Lloyd and L. Maccone, Phys. Rev. A {\bf 67} 052109, (2013).
  
\bibitem{bib:delcampo2013}  A. del Campo, I. L. Egusquiza, M. B. Plenio, and S. F. Huelga, Phys. Rev. Lett. \textbf{110}, 050403 (2013).
  
\bibitem{bib:davidovich2013}  M. M. Taddei, B. M. Escher, L. Davidovich, and R. L. de Matos Filho, Phys. Rev. Lett. \textbf{110}, 050402 (2013).
 
\bibitem{bib:lutz2013}  S. Deffner and E. Lutz, Phys. Rev. Lett. \textbf{111}, 010402 (2013).
  
\bibitem{bib:deffner2013}  S. Deffner and E. Lutz,  J. Phys. A: Math. Theor. \textbf{46} 335302 (2013).

\bibitem{bib:nos_qsl2013} P.M. Poggi, F.C. Lombardo and D.W. Wisniacki, Europhys. Lett. {\bf 104}, 40005 (2013).
  
\bibitem{bib:andersson2014} O. Andersson and N. Heydari, J. Phys. A: Math. and Theo. {\bf 47}, 215301 (2014).


\bibitem{bib:arranz2004} F. J. Arranz, R. M. Benito, and F. Borondo,  J. Chem. Phys. \textbf{120}, 6516 (2004).

\bibitem{bib:nos_molec2014} P. M. Poggi, F. J. Arranz, R. M. Benito, F. Borondo, and D. A. Wisniacki, Phys. Rev. A \textbf{90}, 062108 (2014).

\bibitem{bib:tichy2013} M.C. Tichy \textit{et al.}, Phys. Rev. A {\bf 87}, 063422 (2013).

\bibitem{bib:vliegen2004} E. Vliegen, H.J. Wörner, T.P. Softley and F. Merkt, Phys. Rev. Lett. \textbf{92}, 033005 (2004).

\bibitem{bib:dicarlo2009} L. DiCarlo, et al,  Nature {\bf 460}, 240 (2009)

\bibitem{bib:murgida2007} G.E. Murgida, D.A. Wisniacki and P.I. Tamborenea, Phys. Rev. Lett. \textbf{99}, 036806 (2007).

\bibitem{bib:nos_control2013} P. M. Poggi, F. C. Lombardo and D. A. Wisniacki, Phys. Rev. A \textbf{87}, 022315 (2013).

\bibitem{bib:nos_qsl2015} P.M. Poggi, F.C. Lombardo and D.W. Wisniacki, to appear in  J. Phys. A: Math. and Theo. as a Fast Track Communication (2015).

\bibitem{bib:zener1932} C. Zener, Proc. R. Soc. London, Ser. A \textbf{137}, 696 (1932).
  
\bibitem{bib:nori2010} S.N. Shevchenko, S. Ashhab and F. Nori, Phys. Rep. {\bf 492}, 1-30 (2010).
  
\bibitem{bib:zurek2005} W.H. Zurek, U. Dorner and P. Zoller, Phys. Rev. Lett. {\bf 95}, 105701 (2005).

\bibitem{bib:bason2012} M.G. Bason \textit{et al.}, Nat. Phys. \textbf{8}, 147-152 (2012).

\bibitem{bib:malossi2013} N. Malossi \textit{et al.}, Phys. Rev. A \textbf{87}, 012116 (2013).

\bibitem{bib:hegerfeldt2014} G.C. Hegerfeldt, Phys. Rev. A \textbf{90}, 032110 (2014).

\bibitem{bib:muga2010} I. Lizuain \textit{et al.}, Phys. Rev. A \textbf{82}, 065602 (2010).

\bibitem{bib:sola1999} I. Solá \textit{et al.}, Phys. Rev. A \textbf{59} (6), 4494 (1999).

\bibitem{bib:guerin2003} S. Gu\`erin and H.R. Jauslin, \textit{Control of Quantum Dynamics by Laser Pulses: Adiabatic Floquet Theory, in Advances in Chemical Physics}, Volume 125 (Wiley, Hoboken, 2003)

\bibitem{bib:wisniacki1996} M.J. S\'anchez, E. Vergini, and D.A. Wisniacki, Phys. Rev. E {\bf 54}, 4812 (1996).

\bibitem{bib:werschnik2007} J. Werschnik, and E. Gross, J. Phys. B: At. Mol. Opt. Phys. \textbf{40} R175, (2007).

\bibitem{bib:somloi1993} J. Somlói, V. Kazakov, V. and D.J. Tannor, Chem. Phys. \textbf{172} 85–98. (1993)

\bibitem{bib:montangero2007} S. Montangero, T. Calarco and R. Fazio, Phys. Rev. Lett. \textbf{99} 170501, (2007).

\bibitem{bib:zeilinger1994} M. Reck, A. Zeilinger, H. J. Bernstein, and P. Bertani, Phys. Rev. Lett. \textbf{73}, 58 (1994).

\bibitem{bib:nori2007} S. Ashhab, J. R. Johansson, A. M. Zagoskin, and F. Nori, Phys. Rev. A \textbf{75}, 063414 (2007).

\bibitem{bib:creffield2014} C.E. Creffield and F. Sols, Phys. Rev. A 90, 023636 (2014).

%

}
\end{thebibliography}
\end{document}